\documentclass[lettersize,journal]{IEEEtran}
\usepackage{amssymb,amsmath}
\usepackage{cite}
\usepackage{graphicx}
\usepackage{psfrag}
\usepackage{url}
\usepackage[T1]{fontenc}
\usepackage[utf8]{inputenc}
\usepackage[absolute,overlay]{textpos}
\usepackage{gensymb}
\usepackage{bbm}
\usepackage{cases}
\usepackage[font=footnotesize]{caption}
\usepackage[font=footnotesize]{subcaption}
\usepackage{float}
\usepackage[linesnumbered,ruled,lined]{algorithm2e}
\SetKwInOut{Training}{Train}
\SetKwInOut{Testing}{Test}
\usepackage{pgf}

\usepackage[nocomma]{optidef}

\usepackage{amsthm}

\usepackage{booktabs}
\usepackage{color,soul}

\usepackage{textcomp}
\usepackage{xcolor}
\def\BibTeX{{\rm B\kern-.05em{\sc i\kern-.025em b}\kern-.08em
    T\kern-.1667em\lower.7ex\hbox{E}\kern-.125emX}}

\usepackage{tabularx}
\usepackage{makecell}
\usepackage{multirow}

\begin{document}
\title{Meta-Offline and Distributional Multi-Agent RL for Risk-Aware Decision-Making}
\author{
	\IEEEauthorblockN{Eslam Eldeeb and Hirley Alves \\
	}
	\IEEEauthorblockA{Centre for Wireless Communications (CWC), University of Oulu, Finland \\
	Email: firstname.lastname@oulu.fi}
}
\maketitle

\begin{abstract}
Mission critical applications, such as UAV-assisted IoT networks require risk-aware decision-making under dynamic topologies and uncertain channels. We propose meta-conservative quantile regression (M-CQR), a meta-offline distributional MARL algorithm that integrates conservative Q-learning (CQL) for safe offline learning, quantile regression DQN (QR-DQN) for risk-sensitive value estimation, and model-agnostic meta-learning (MAML) for rapid adaptation. Two variants are developed: meta-independent CQR (M-I-CQR) and meta-CTDE-CQR. In a UAV-based communication scenario, M-CTDE-CQR achieves up to $50 \%$ faster convergence and outperforms baseline MARL methods, offering improved scalability, robustness, and adaptability for risk-sensitive decision-making. Code is available at \url{https://github.com/Eslam211/MA_Meta_ODRL}
\end{abstract}

\vspace{-2mm}
\section{Introduction}\label{sec:introduction}
Unmanned aerial vehicles (UAVs) are increasingly deployed in environments where safety and regulatory constraints impose strict “no-fly” and high-hazard zones. In such settings, ranging from urban canyons to wildfire-prone regions, \textit{risk sensitive decision-making}, for instance, in the form of risk-aware trajectory planning, \textit{is not merely desirable but mission-critical}, as violations can result in catastrophic loss of assets, civilian harm, or regulatory penalties \cite{10485273}. 

Deep reinforcement learning (RL) and multi-agent RL (MARL) methods, such as deep Q-networks (DQNs), have been widely applied to complex decision-making problems~\cite{10829755,10021988,10448459,10095748}. However, MARL faces three main challenges: \textit{i)} reliance on online interactions, which can be unsafe or impractical~\cite{levine2020offline,10078377}; \textit{ii)} optimisation based on average behaviour, ignoring real-world uncertainty~\cite{bdr2023}; and \textit{iii)} re-optimisation from scratch when network settings or objectives change, incurring high computational cost~\cite{9495238}.

Offline reinforcement learning (RL) enables policy learning from pre-collected datasets, avoiding unsafe online interactions. Conservative Q-learning (CQL)~\cite{kumar2020conservative} adapts MARL by adding a conservative term to the Bellman update, improving performance with fixed offline data. However, offline RL must still address distributional shift, which can cause over-optimistic value estimates in unseen states~\cite{10888196}.

Distributional RL models the full return distribution, enabling direct optimization of risk measures such as CVaR. Techniques like quantile regression DQN (QR-DQN)~\cite{dabney2017distributional} estimate Q-value distributions via quantile regression. Complementary to offline and distributional MARL, model-agnostic meta-learning (MAML)~\cite{FINN} leverages knowledge across tasks to initialize parameters for rapid adaptation to unseen tasks.


Several studies have explored offline MARL, distributional MARL, and MAML in different domains. For example,~\cite{10753476} combines CQL and QR-DQN for UAV trajectory planning but does not address retraining under changing network configurations. In~\cite{9410247}, an online multi-agent meta-RL algorithm based on MAPPO is proposed for adaptive multipath routing but still risks unsafe exploration. Similarly,~\cite{10530871} develops a meta-offline single-agent RL method for high-speed railway control.


A key research gap remains in unifying offline and distributional MARL with meta-learning for safe, risk-sensitive, and adaptive decision-making. We propose Meta-Conservative Quantile Regression (M-CQR), which integrates Conservative Q-Learning (CQL) for safe offline evaluation, Quantile-Regression DQN (QR-DQN) for risk-sensitive value estimation, and MAML for rapid task adaptation. We demonstrate M-CQR on risk-aware UAV trajectory planning in hazardous areas.

The contributions of this work are:
\begin{itemize}
    \item We present two unified frameworks, meta-independent-CQR (M-I-CQR) and meta-centralized training decentralized execution-CQL (M-CTDE-CQR), combining offline MARL (CQL), distributional MARL (QR-DQN), and meta-learning (MAML) for real-world decision-making.
    

    \item We evaluate the designed algorithm in UAV trajectory and scheduling optimization to jointly maximize information freshness and power efficiency in a dynamic, risk-prone environment using static offline data.
    
    \item Simulation results showing centralized MARL outperforms independent MARL, and both meta-MARL variants outperform all baselines.
\end{itemize}

This paper is organized as follows: Section~\ref{sec:Preliminaries} introduces the preliminaries, Section \ref{sec:off-m-cql} describes the proposed approach, Section \ref{sec:results} discusses the numerical results on a selected UAV application, and Section \ref{sec:conclusions} concludes the paper.

\vspace{-2mm}
\section{Background}\label{sec:Preliminaries}

\subsection{Partially Observable Markov Decision Process}
Multi-agent environments with $I$ agents are commonly modeled as partially observable Markov decision processes (POMDPs), defined by the tuple $\langle S,A,P,R,O,\gamma \rangle$, where $S$ is the state space, $A$ the action space, $P$ the transition probabilities, $R$ the reward function, $O$ the observation space, and $\gamma$ the discount factor. Each agent $i$ observes ${o}^i$, takes action ${a}^i$, receives reward $r$, and transitions to ${o}^{\prime i}$ with probability $p({o}^{\prime i}|{o}^i,{a}^i)$. In a cooperative setting, the goal is to maximize the expected discounted return $\mathbb{E}\bigg[\sum_{t=0}^{\infty}\gamma^t r_t\bigg]$, where $r_t$ is the reward at time $t$.

For independent training, the DQN loss for each agent is:
\begin{align}
\label{bellman_error}
    \mathcal{L}_{\text{I-DQN}}^i \!=& \hat{\mathbb{E}} \left[ \!\left(\!  r \!+\!\gamma \max_{a^{\prime i}} \hat{Q}^{i(k)}(o^{\prime i},a^{\prime i}) 
   \!-\! Q^i(o^i,a^i) \!\right)^2 \!\right],
\end{align}
where $\hat{\mathbb{E}}[\cdot]$ is the sample mean over experiences from the offline dataset $\mathcal{B}$, $a^{\prime i}$ is the next action and $\hat{Q}^{i(k)}$ is the current estimate of the optimal Q-function for agent $i$ at iteration $k$. The Q-function $Q(o^i,a^i)$ is typically parameterized by a neural network.

\vspace{-2mm}
\subsection{Conservative Q-Learning}
Directly applying DQN to offline datasets fails due to out-of-distribution (OOD) shift between dataset and learned policies. Conservative Q-learning (CQL) mitigates this by adding a conservative term to the DQN loss. For agent $i$, the CQL loss is
\vspace{-1mm}
\begin{align}
\label{Ind_CQL_MARL}
    &\mathcal{L}_{\text{I-CQL}}^i = \frac{1}{2}\mathcal{L}_{\text{I-DQN}}^i 
    \nonumber \\ 
    &\quad
    + \alpha \hat{\mathbb{E}}\! \bigg[\!\log\!\! \bigg(\!\! \sum_{\tilde a^i} 
    \exp (\!Q^i(o^i,\tilde{a}^i)\! )\!\bigg)
    \!\!-\! Q^i(o^i, a^i)\! \bigg],
\end{align}
where $\tilde{a}^i$ enumerates all actions and $\alpha > 0$ controls conservatism~\cite{pan2022plan}. The resulting algorithm is termed independent-CQL (I-CQL).

Another approach is centralized training with decentralized execution (CTDE), where a global Q-function is estimated via value decomposition~\cite{sunehag2017value}
\vspace{-1mm}
\begin{align}
\label{value_dec_eq}
   Q(s,a) = \sum_{i=1}^I \tilde{Q}^i(o^i,a^i),
\end{align}
with $s$ and $a$ denoting joint state and action spaces. The CTDE DQN loss becomes
\vspace{-1mm}
\begin{align}
\label{Ind_DQN_CTDE}
\mathcal{L}_{\text{CTDE-DQN}}\! = \hat{\mathbb{E}}\! \Bigg[\!\! \Bigg(\! r \!+\! \gamma \sum_{i=1}^I\! \max_{\tilde{a}^{i}} \hat{Q}^{i(k)} \! (o^{\prime i},\tilde{a}^i) \!-\!\sum_{i=1}^I \tilde{Q}^i(o^i,a^i)\!\!\Bigg)^{\!\!2} \Bigg].
\end{align}
Similarly, we define a single CQL loss to be used by all agents
\vspace{-1mm}
\begin{align}
\label{Ind_CQL_CTDE} 
  &\mathcal{L}_{\text{CTDE-CQL}} \!=\! \frac{1}{2} \mathcal{L}_{\text{CTDE-DQN}} \nonumber \\
  &\!+\! \alpha \, \hat{\mathbb{E}}\! \sum_{i=1}^I \!\bigg[ \!\log \!\bigg(\! \sum_{\tilde a^i} \exp ( \tilde{Q}^i(o^i,\tilde{a}^i) ) \!\bigg)  \!-\! \tilde{Q}^i(o^i, a^i) \!\bigg].
\end{align}
We refer to this algorithm as a centralised training, decentralised execution-CQL (CTDE-CQL) algorithm.

\subsection{Quantile-Regression DQN}
Quantile-regression DQN (QR-DQN) approximates the return distribution via quantile regression, enabling risk-sensitive policies based on measures such as conditional value-at-risk (CVaR) rather than mean return. It minimizes the Wasserstein distance using the quantile regression loss~\cite{dabney2017distributional}
\begin{equation}
\label{QRDQN_eq}
    \mathcal{L}_{\text{QR-DQN}} = \frac{1}{N^2} \sum_{j=1}^N \sum_{j^{\prime}=1}^N \zeta_{\tau}(r + \gamma Z_{j^{\prime}}(o^{\prime},a^{\prime}) - Z_j(o,a)),
\end{equation}
where $Z_j$ and $Z_{j^{\prime}}$ are predicted and target quantiles, respectively. Their difference corresponds to the temporal-difference error in the Bellman update~\cite{dabney2017distributional}. The Huber quantile loss is
\begin{align}
\label{quantile_loss}
   \zeta_{\tau}^{}(u) &= \begin{cases}-\frac{1}{2} u^2\left|\tau - \mathbbm{1} \{ u < 0 \}\right|, \quad \quad \quad
    &\text{if } | u |\leq 1, \\
    \left( | u | - \frac{1}{2}  \right)\left|\tau - \mathbbm{1} \{ u < 0 \}\right|,  \quad 
    &\text{otherwise};
   \end{cases}
\end{align}
where $\tau$ denotes fixed probability levels. Selecting lower quantiles focuses optimization on the worst $\tau \%$ of returns, penalizing catastrophic trajectories and promoting safety under uncertainty.


\vspace{-2mm}
\subsection{Model-Agnostic Meta-Learning}
Model-agnostic meta-learning (MAML) enables a parametric model to quickly adapt to unseen tasks by leveraging experience across tasks. Let $\mathcal{T} = {T_1, \cdots, T_U}$ denote $U$ tasks sampled from a distribution $p(T)$. MAML performs inner and outer updates on model parameters $\theta$. The inner update adapts to task $T_u$:~\cite{FINN}
\begin{equation}
\label{maml_1}
    \theta_u^{\prime} = \theta - \eta_{1} \nabla_{\theta}\mathcal{L}({T_u};\theta),
\end{equation}
where $\eta_{1}$ is the inner learning rate. The outer update then adjusts $\theta$ to improve performance across tasks:
\begin{equation}
\label{maml_2}
    \theta \leftarrow \theta - \eta_{2} \nabla_{\theta}\sum_{u=1}\mathcal{L}({T_u};\theta_u^{\prime}),
\end{equation}
where $\eta_{2}$ is the outer learning rate.

\vspace{-2mm}
\section{The Offline Meta-CQR Framework}\label{sec:off-m-cql} 

We propose Meta-CQR, a meta-offline MARL framework that integrates CQL, QR-DQN, and MAML under two training paradigms: \emph{independent} and \emph{centralized training with decentralized execution (CTDE)}.

For independent training, we define the I-CQR loss as a combination of QR-DQN’s quantile regression loss and CQL’s conservative regularization
\begin{align}
\label{Ind_CQR_revised}
&\mathcal{L}_{\text{I-CQR}}^i = \frac{1}{N^2} \hat{\mathbb{E}} \sum_{j=1}^N \sum_{j^{\prime}=1}^N \zeta_{\tau}\bigg(r + \gamma Z^{i (k)}_{j^{\prime}}(o^{i \prime},a^{i \prime}) - Z^i_j(o^i,a^i)\bigg) \nonumber \\ &
\,\, \!+\! \alpha \hat{\mathbb{E}}\Biggl[\! \frac{1}{N} \sum_{j=1}^N \Biggl[\!\log \!\sum_{\tilde{a}^i}
\exp \bigl( Z_j^i(o^i,\tilde{a}^i) \bigr) \!-\! Z_j^i(o^i,a^i)\! \Biggr]\! \Biggr], 
\end{align}
where $Z_j^i(o^i,a^i)$ is the predicted $j^{\text{th}}$ quantile for agent $i$ and $Z_{j^{\prime}}^{i (k)}(o^{i \prime},a^{i \prime})$ is the next-state target quantile.


Extending I-CQR to the CTDE setting yields the CTDE-CQR loss
\begin{align}
\label{CTDE_CQR}
&\mathcal{L}_{\text{CTDE-CQR}} = \\ \nonumber
&\frac{1}{N^2} \hat{\mathbb{E}} \sum_{j=1}^N \sum_{j^{\prime}=1}^N \zeta_{\tau}\bigg(r + \gamma \sum_{i=1}^I \tilde{Z}^{i (k)}_{j^{\prime}}(o^{i \prime},a^{i \prime}) - \sum_{i=1}^I \tilde{Z}^i_j(o^i,a^i)\bigg) \\&
+ \alpha \hat{\mathbb{E}} \sum_{i=1}^I \Biggl[ \frac{1}{N} \sum_{j=1}^N \Biggl[\log \sum_{\tilde{a}^i}
\exp \biggl( \tilde{Z}_j^i(o^i,\tilde{a}^i) \biggr) - \tilde{Z}_j^i(o^i,a^i) \Biggr] \Biggr], \nonumber
\end{align}
where $Z^i$ is rewritten as $\tilde{Z}^i$ due to the value-decomposition approximation. 

\begin{algorithm}[!t]
\SetAlgoLined

\textbf{Define} the hyperparameters $D$, $U$, $\gamma$, $\eta_{1}$, $\eta_{2}$, $\alpha$, $N$, $U$ task distribution $p(T)$, and training epochs $E_{\text{meta}}$

Initialize the Q-networks initial parameters $\{\theta^i\}_{i=1}^I$

Collect an offline dataset $\{\mathcal{B}\}_{i=1}^I$ for each environment and divide it into support and query sets.

\For{\text{epochs} $e$ in $\{1, \cdots,E_{\text{meta}}\}$}{


\For{\text{task} in $\{T_1, \cdots,T_U\}$}{

\If{\textsc{M-CTDE-CQR}}{
Estimate the global Q-function using value decomposition as in~\eqref{value_dec_eq}
}

\For{\text{agent} in $\{1, \cdots,I\}$}{
Update the initial weights $\theta^i$ using the support set as in~\eqref{task_update} using $\mathcal{L}_{\text{I-CQL}}^i(T_u;\theta^i)$ or $\mathcal{L}_{\text{CTDE-CQL}}(T_u;\theta^i)$

}
}
\For{\text{agent} in $\{1, \cdots,I\}$}{
Using the query set and the updated parameters, calculate the meta-losses $\mathcal{L}_{\text{meta}}$ using $\mathcal{L}_{\text{I-CQL}}^i(T_u;\theta_u^{\prime i})$ or $\mathcal{L}_{\text{CTDE-CQL}}(T_u;\theta_u^{\prime i})$ and update the initial weights $\theta^i$ using~\eqref{meta_optimization}
}
}

\textbf{Return} Q-networks initial parameters $\{\theta^i\}_{i=1}^I$

\caption{The proposed M-I-CQR and M-CTDE-CQR algorithms.}
\label{Meta_off_Alg}
\end{algorithm}

To design adaptive I-CQR and CTDE-CQR algorithms, we combine them with MAML, where the objective is to find the initial parameters $\theta^i$ for each agent $i$ that rapidly adapts to new tasks in a few stochastic gradient descent (SGD) steps. In MAML, the offline dataset $\mathcal{B}$ is split into offline support set $\mathcal{B}_{\text{support}}$ and offline query set $\mathcal{B}_{\text{query}}$. To integrate the MAML algorithm into the proposed I-CQR and CTDE-CQR algorithms, we can easily rewrite~\eqref{maml_1} and~\eqref{maml_2} by replacing the losses with algorithm-specific losses. For the independent CQR case, each agent $i$ updates the initial parameters for each task $\tau_i$ using the offline support set $\mathcal{B}_{\text{support}}$ as follows
\begin{equation}
    \label{task_update}
    \theta_u^{\prime i} \leftarrow \theta^i - \eta_{1}\nabla_{\theta^i} \mathcal{L}_{\text{I-CQR}}^i(T_u;\theta^i),
\end{equation}
where $\theta^i$ is the Q-network parameters of agent $i$, $\theta_u^{\prime i}$ is the updated parameters, $\eta_{\text{inner}}$ is a learning rate and $\mathcal{L}_{\text{CQL}}^u(T_u;\theta^i)$ is the CQL loss of agent $i$ in an environment that corresponds to task $\tau_u$ using parameters $\theta^i$. Afterwards, meta-losses are calculated for each agent by summing the losses across all tasks using the new task-specific parameters applied to the offline query set $\mathcal{B}_{\text{query}}$ as follows
\begin{equation}
    \label{meta_optimization}
    \theta^i \leftarrow \theta^i - \eta_{2} \nabla_{\theta^i}\sum_{u=1}\mathcal{L}_{\text{I-CQR}}^i(T_u;\theta_u^{\prime i}).
\end{equation}
The same procedure is applied to the CTDE CQR case by replacing $\mathcal{L}_{\text{I-CQR}}^i(T_u;\theta^i)$ and $\mathcal{L}_{\text{I-CQR}}^i(T_u;\theta_u^{\prime i})$ to $\mathcal{L}_{\text{I-CTDE}}(T_u;\theta^i)$ and $\mathcal{L}_{\text{I-CTDE}}(T_u;\theta_u^{\prime i})$ in equations \eqref{task_update} and \eqref{meta_optimization}, respectively.
\textbf{Algorithm \ref{Meta_off_Alg}} summarizes the proposed meta-independent-CQR (M-I-CQR) and meta-CTDE-CQR (M-CTDE-CQR) approaches.

\begin{figure*}[t!]
    \centering
    \subfloat[Independent\label{Indep}]{\includegraphics[width=0.65\columnwidth]{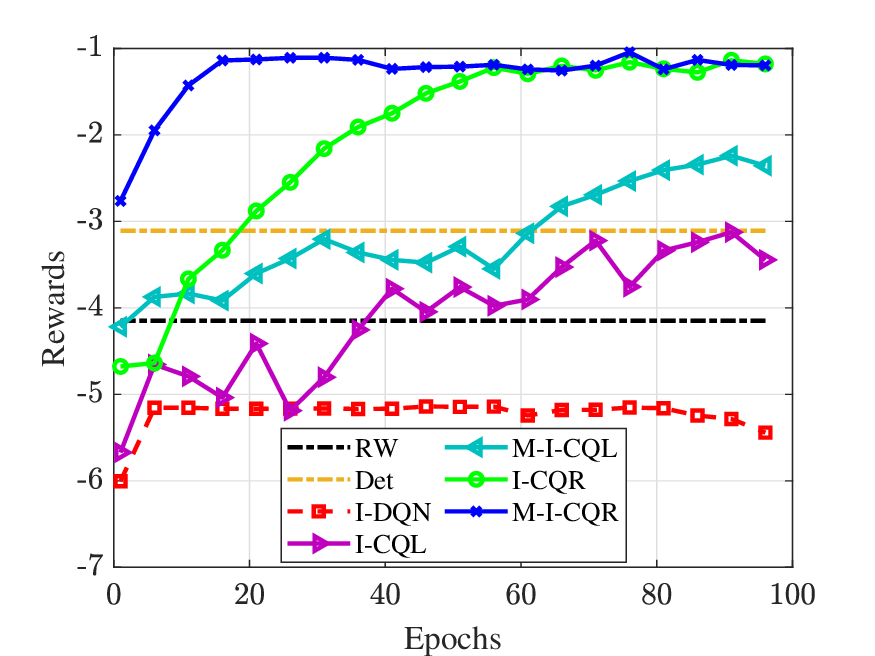}}
    \subfloat[CTDE\label{CTDE}]{\includegraphics[width=0.65\columnwidth]{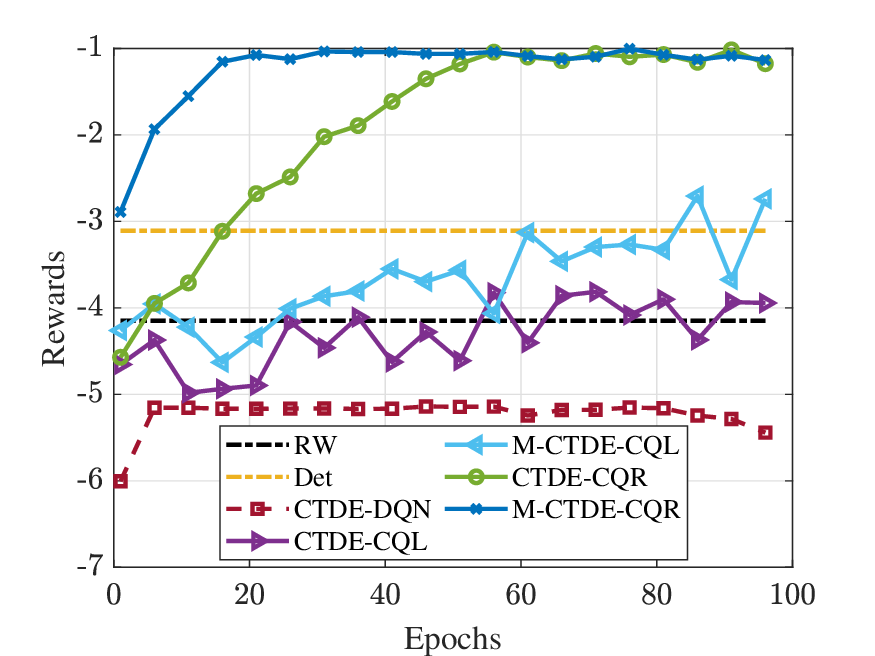}}
    \vspace{-1mm}
    \caption{Normalised rewards convergence of the proposed algorithm compared to the benchmarks: (a) independent training case and (b) CTDE training case.}
    \vspace{-2mm}
    \label{Conv}
\end{figure*}
\begin{figure*}[h!]
    \centering
    \subfloat[Shots\label{Shots}]{\includegraphics[width=0.65\columnwidth]{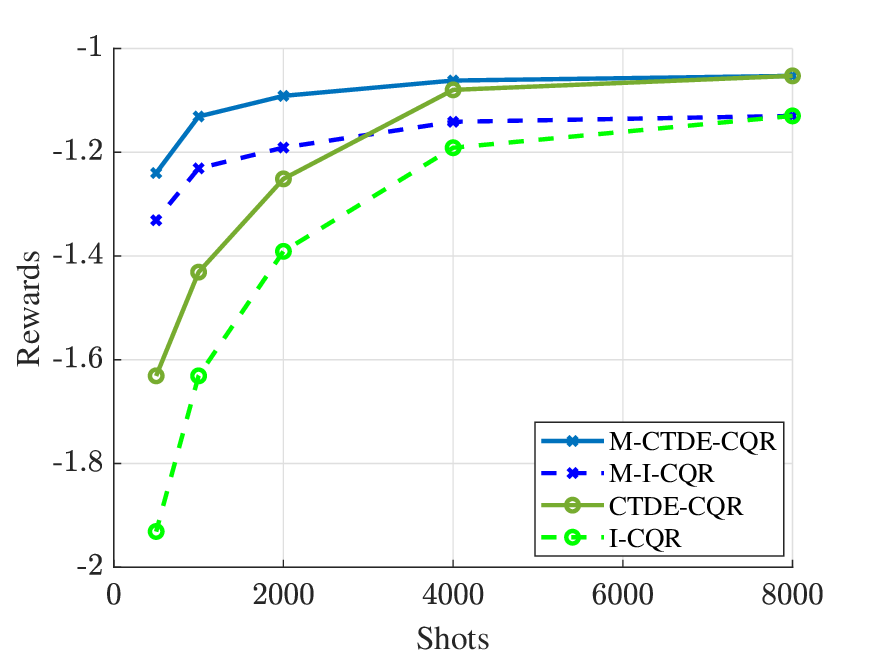}}
    \subfloat[Tasks\label{Tasks}]{\includegraphics[width=0.65\columnwidth]{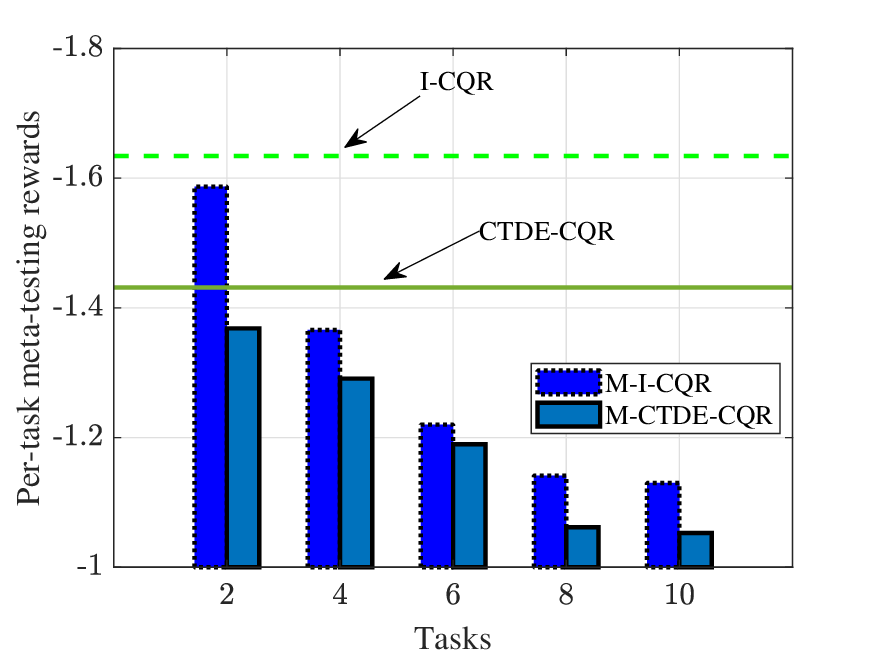}}
    \subfloat[AoI and power trade-off\label{AoI_Pwr}]{\includegraphics[width=0.65\columnwidth]{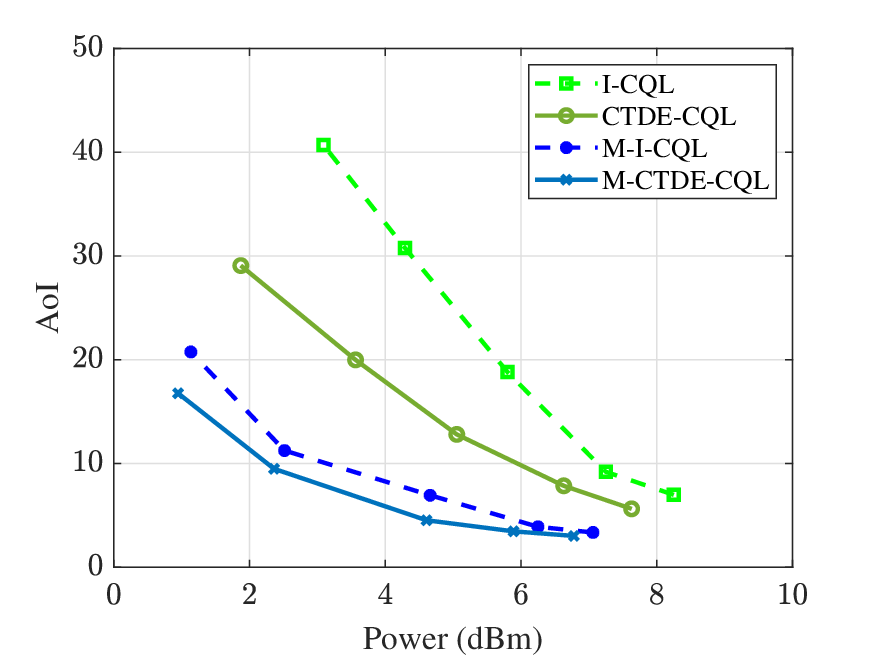}}
    \vspace{-1mm}
    \caption{The effect of model parameters on testing: (a) dataset size, (b) training tasks, and (c) changing $\lambda$.}
    \vspace{-2mm}
    \label{Testing}
\end{figure*}

\vspace{-2mm}
\section{Numerical Results and Discussion}\label{sec:results}
%
\subsection{Environment Description}
We consider an $L \times L$ grid world with $D$ ground IoT devices $\mathcal{D}=\{1,\dots,D\}$ uniformly placed at cell centers. Each device $d$ has coordinates $(x^d,y^d)$ and is served by $I$ rotary-wing UAVs $\mathcal{I}=\{1,\dots,I\}$, each flying at altitude $h_i$. 

The environment is episodic with discrete time steps. At time $t$, UAV $i$ is at $(x^i_t,y^i_t)$ on a 2D plane. In one time step, a UAV either flies distance $r_l$ (between adjacent cells) with velocity $v^i$ or hovers to receive a device uplink packet. We assume Rayleigh fading and optimize adaptive transmission power $P_t^d$ per channel state. The AoI (Age of Information) of device $d$ evolves as $A^d_t = A^d_{t-1} + 1$ if unserved, and resets to $A^d_t=1$ when served~\cite{AoI_orig}. The system model and parameters follow~\cite{10753476,eldeeb2022multi}.

\noindent\textbf{Problem formulation:} The main objective is to determine the optimum trajectories of the UAVs and their scheduling policies to minimise the AoI and devices' transmit power jointly. This problem is formulated as a POMDP as follows
\begin{enumerate}
    \item \textbf{Observation} $o_t^i$: At time $t$, each UAV $i$ observes its position $(x^i_t,y^i_t)$ and the AoI of the devices $(A^1_t, A^2_t, \!\cdots\!, A^D_t)$. Hence, $o_t^i \!=\! (x^i_t,y^i_t,A^1_t, A^2_t, \!\cdots\!, A^D_t)$ and the total state space of the system is $s_t \!=\! (x^1_t, y^1_t, \!\cdots\!, x^I_t, y^I_t, A^1_t, A^2_t, \!\cdots\!, A^D_t)$.

    \item \textbf{Action} $a_t^i$: At time $t$, the action space of UAV $i$ is $a_t^i = (w^i_t,s^i_t)$, where $w^i_t \!=\! \{\text{east}, \text{west},\text{north},\text{south},\text{hover}\}$ is the movement direction and $s^i_t \!=\! d$ is the scheduled device.

    \item \textbf{Reward} $r_t$: We consider a cooperative MARL problem, where all agents cooperate towards a single objective. Hence, the reward is formulated as ($r_t = - \sum_{d = 1}^{D}\delta_d A^d_t - \lambda P^d_t - pen$), 
    where $\delta_d$ is a weight factor representing the importance of device $d$ and $\lambda$ is a scaling variable controlling the trade-off between the AoI and the transmission power. We assume a risk region in the grid world, where the condition "if risk" means that one of the UAVs enters the risk region and receives a heavy bad reward ($pen$) with a small probability $p_{\text{risk}}$.
\end{enumerate}

We consider a relatively small-size static offline dataset $\mathcal{B}$, which contains some experiences $\langle \{o_t^i\}_{i=1}^{I},\{a_t^i\}_{i=1}^{I},\{o_{t+1}^i\}_{i=1}^{I},r_t \rangle$ collected using behavioral policies, without access to any online interaction with the environment. In addition, we consider each environment as a single task with unique device positions, risk region coordinates, and $\lambda$ values. Therefore, we aim to develop a scalable algorithm that can quickly adapt to new, unseen environments.

\vspace{-2mm}
\subsection{Simulation Results}

We evaluate the proposed M-I-CQR and M-CTDE-CQR algorithms against two benchmarks: random walk (RW), where agents act randomly, and deterministic (det), where agents follow shortest paths to cover different network areas. An ablation study compares our methods to I-DQN, CTDE-DQN, I-CQL, CTDE-CQL, M-I-CQL, M-CTDE-CQL, I-CQR, and CTDE-CQR. Simulations are conducted in a $1000$ m $\times$ $1000$ m network with $D=10$ devices and $I=2$ UAVs. The Q-network uses two hidden layers of $256$ neurons each. Offline datasets are the final $10 \%$ of an online DQN agent’s experience. Experiments run on a single NVIDIA Tesla V100 GPU with PyTorch.

\begin{table}[t!]
\centering
 \caption{Performance evaluation over $100$ unseen test environments (each tested online $100$ times after $20$ fine-tune offline training iterations).
}
\label{tab_perf}
\begin{tabular}{@{}cccccc@{}}
\toprule
\textbf{Alg.} & $\:$ Avg. return $\:$ & $\:$ $\text{CVaR}$ $\:$ & AoI $\:$ & $\:$ Pwr (dBm) $\:$ & $\:$ Viol. $\:$ \\ \midrule
\midrule
I-CQR & $-2.4$ & $-2.9$ & $18.845$ & $5.94$ & $13.34\%$ \\
\hline
M-I-CQR & $-1.8$ & $-2.1$ & $10.87$ & $3.79$ & $4.24\%$ \\
\hline 
CTDE-CQR & $-2.3$ & $-2.6$ & $16.26$ & $4.89$ & $12.20\%$ \\
\hline 
M-CTDE-CQR & $\boldsymbol{-1.7}$ & $\boldsymbol{-2.1}$ & $\boldsymbol{8.32}$ & $\boldsymbol{2.87}$ & $\boldsymbol{3.79\%}$ \\
\bottomrule
\end{tabular}
\end{table}

Fig.~\ref{Conv} shows reward convergence during meta-testing, trained on $10$ tasks with a $5000$-entry offline dataset, compared to independent and CTDE benchmarks. I-DQN and CTDE-DQN fail due to distributional shift. M-I-CQR and M-CTDE-CQR converge within $20$ epochs and achieve the highest rewards, while their non-MAML variants (I-CQR, CTDE-CQR) require over $50$ epochs. Offline methods without distributional RL (I-CQL, M-I-CQL, CTDE-CQL, M-CTDE-CQL) exhibit high variance and fail to reach optimal behavior.


Fig.~\ref{Testing} shows the effect of hyperparameters on performance with $50$ training epochs. Fig.~\ref{Shots} examines dataset size (shots) with $10$ tasks, where larger datasets yield higher, more stable rewards. Fig.~\ref{Tasks} demonstrates that more training tasks improve Q-network initialization and rewards, with even $2$ tasks outperforming random initialization. Fig.~\ref{AoI_Pwr} reports achievable AoI and transmission power for agents trained with and without MAML. M-I-CQR and M-CTDE-CQR consistently achieve higher rewards and lower AoI/power, with CTDE providing superior stability due to Q-network sharing among agents. 

Table~\ref{tab_perf} reports inference results on $100$ unseen environments using CVaR ($15 \%$) to capture worst-case returns. M-CTDE-CQR achieves the highest average and CVaR returns, lowest AoI and transmission power, and only $3.79 \%$ risk-region violations versus $12.20 \%$ for CTDE-CQR, demonstrating the benefit of distributional RL for risk avoidance. From the ablation comparisons in Fig.~\ref{Conv}, Fig.~\ref{Testing}, and Table~\ref{tab_perf}, distributional RL (CQR) mainly reduces risk-region violations and improves CVaR, meta-learning (MAML) accelerates adaptation and improves worst-case performance, while CTDE enhances stability via value decomposition.

\section{Conclusions}\label{sec:conclusions} 
%
%
This paper presented a multi-agent meta-offline and distributional RL framework (M-CQR) for risk-sensitive decision-making. By combining CQL and QR-DQN, M-CQR enables safe training from static offline datasets while accounting for risk. Integrating CQR with MAML allows rapid adaptation to changing network objectives. Two variants were developed: M-I-CQR (independent training) and M-CTDE-CQR (CTDE). UAV simulations show that M-CTDE-CQR converges faster and more stably than M-I-CQR, with both outperforming offline MARL baselines. M-CTDE-CQR achieves up to $50 \%$ faster adaptation. This makes the approach particularly suitable for mission-critical applications such as search-and-rescue or wildfire monitoring.

\vspace{-1mm}
\section*{Acknowledgments} \vspace{1mm}
This work was supported by 6G Flagship (Grant Number 369116) funded by the Research Council of Finland.

\vspace{-1mm}

\bibliographystyle{IEEEtran}
\bibliography{IEEEabrv,references}
\end{document}